\documentclass[twoside,twocolumn,9pt]{article}
\usepackage{extsizes}
\usepackage[super,sort&compress,comma]{natbib} 
\usepackage[version=3]{mhchem}
\usepackage[left=1.5cm, right=1.5cm, top=1.785cm, bottom=2.0cm]{geometry}
\usepackage{balance}
\usepackage{mathptmx}
\usepackage{sectsty}
\usepackage{graphicx} 
\usepackage{lastpage}
\usepackage{upgreek}
\usepackage{amsmath} 
\usepackage{gensymb}
\usepackage[format=plain,justification=justified,singlelinecheck=false,font={stretch=1.125,small,sf},labelfont=bf,labelsep=space]{caption}
\usepackage{float}
\usepackage{fancyhdr}
\usepackage{fnpos}
\usepackage[english]{babel}
\addto{\captionsenglish}{%
  
}
\usepackage{array}
\usepackage{droidsans}
\usepackage{charter}
\usepackage[T1]{fontenc}
\usepackage[usenames,dvipsnames]{xcolor}
\usepackage{setspace}
\usepackage[compact]{titlesec}
\usepackage{hyperref}

\usepackage{epstopdf}

\usepackage{color,ulem}

\definecolor{cream}{RGB}{222,217,201}

\begin{document}

\pagestyle{fancy}
\thispagestyle{plain}
\fancypagestyle{plain}{
\renewcommand{\headrulewidth}{0pt}
}

\makeFNbottom
\makeatletter
\renewcommand\LARGE{\@setfontsize\LARGE{15pt}{17}}
\renewcommand\Large{\@setfontsize\Large{12pt}{14}}
\renewcommand\large{\@setfontsize\large{10pt}{12}}
\renewcommand\footnotesize{\@setfontsize\footnotesize{7pt}{10}}
\makeatother

\renewcommand{\thefootnote}{\fnsymbol{footnote}}
\renewcommand\footnoterule{\vspace*{1pt}%
\color{cream}\hrule width 3.5in height 0.4pt \color{black}\vspace*{5pt}} 
\setcounter{secnumdepth}{5}

\makeatletter 
\renewcommand\@biblabel[1]{#1}            
\renewcommand\@makefntext[1]%
{\noindent\makebox[0pt][r]{\@thefnmark\,}#1}
\makeatother 
\renewcommand{\figurename}{\small{Fig.}~}
\sectionfont{\sffamily\Large}
\subsectionfont{\normalsize}
\subsubsectionfont{\bf}
\setstretch{1.125} 
\setlength{\skip\footins}{0.8cm}
\setlength{\footnotesep}{0.25cm}
\setlength{\jot}{10pt}
\titlespacing*{\section}{0pt}{4pt}{4pt}
\titlespacing*{\subsection}{0pt}{15pt}{1pt}

\fancyfoot{}
\fancyfoot[LO,RE]{\vspace{-7.1pt}\includegraphics[height=9pt]{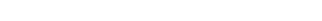}}
\fancyfoot[CO]{\vspace{-7.1pt}\hspace{13.2cm}\includegraphics{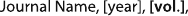}}
\fancyfoot[CE]{\vspace{-7.2pt}\hspace{-14.2cm}\includegraphics{head_foot/RF}}
\fancyfoot[RO]{\footnotesize{\sffamily{1--\pageref{LastPage} ~\textbar  \hspace{2pt}\thepage}}}
\fancyfoot[LE]{\footnotesize{\sffamily{\thepage~\textbar\hspace{3.45cm} 1--\pageref{LastPage}}}}
\fancyhead{}
\renewcommand{\headrulewidth}{0pt} 
\renewcommand{\footrulewidth}{0pt}
\setlength{\arrayrulewidth}{1pt}
\setlength{\columnsep}{6.5mm}
\setlength\bibsep{1pt}

\newcommand{\dlt}[1]{}
\newcommand{\del}[1]{}
\newcommand{\change}[1]{{\color{orange} #1}}
\newcommand{\new}[1]{#1}

\makeatletter 
\newlength{\figrulesep} 
\setlength{\figrulesep}{0.5\textfloatsep} 

\newcommand{\topfigrule}{\vspace*{-1pt}%
\noindent{\color{cream}\rule[-\figrulesep]{\columnwidth}{1.5pt}} }

\newcommand{\botfigrule}{\vspace*{-2pt}%
\noindent{\color{cream}\rule[\figrulesep]{\columnwidth}{1.5pt}} }

\newcommand{\dblfigrule}{\vspace*{-1pt}%
\noindent{\color{cream}\rule[-\figrulesep]{\textwidth}{1.5pt}} }

\makeatother

\twocolumn[
  \begin{@twocolumnfalse}
{\includegraphics[height=30pt]{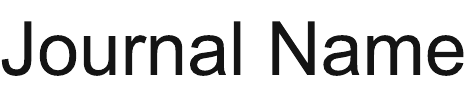}\hfill\raisebox{0pt}[0pt][0pt]{\includegraphics[height=55pt]{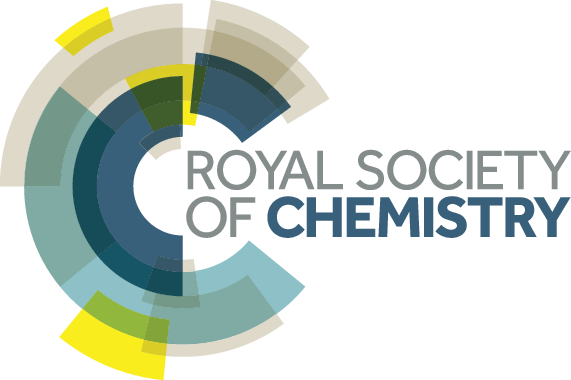}}\\[1ex]
\includegraphics[width=18.5cm]{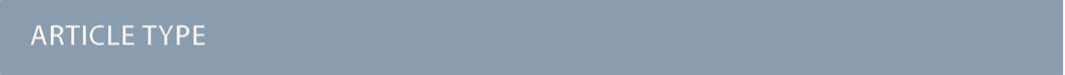}}\par
\vspace{1em}
\sffamily
\begin{tabular}{m{4.5cm} p{13.5cm} }

\includegraphics{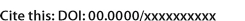} & \noindent\LARGE{\textbf{Analytical electron microscopy analysis of insulating and metallic phases in nanostructured vanadium dioxide}} \\
\vspace{0.3cm} & \vspace{0.3cm} \\

 & \noindent\large{Jan Krpenský,\textit{$^{a}$} Michal Horák,\textit{$^{b}$} Jiří Kabát,\textit{$^{a}$} Jakub Planer,\textit{$^{b}$} Peter Kepič,\textit{$^{b}$} Vlastimil Křápek,$^\ast$\textit{$^{a,b}$} and Andrea Konečná$^\ast$\textit{$^{a,b}$}} \\

\includegraphics{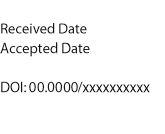} & \noindent\normalsize{Vanadium dioxide (VO$_2$) is a strongly-correlated material that exhibits insulator-to-metal transition (IMT) near room temperature, which makes it a promising candidate for applications in nanophotonics or optoelectronics. However, creating VO$_2$ nanostructures with the desired functionality can be challenging due to microscopic inhomogeneities that can significantly impact the local optical and electronic properties. Thin lamellas, produced by focused ion beam milling from a homogeneous layer, provide a useful prototype for studying VO$_2$ at the truly microscopic level using a scanning transmission electron microscope (STEM). High-resolution imaging is used to identify structural inhomogeneities while electron energy-loss spectroscopy (EELS) supported by statistical analysis helps to detect V$_x$O$_y$ stoichiometries with a reduced oxidation number of vanadium at the areas of thickness below 70~nm. On the other hand, the thicker areas are dominated by vanadium dioxide, where the signatures of IMT are detected in both core-loss and low-loss EELS experiments with in-situ heating. The experimental results are interpreted with ab-initio and semi-classical calculations. This work shows that structural inhomogeneities such as pores and cracks present no harm to the desired optical properties of VO$_2$ samples.} \\

\end{tabular}

 \end{@twocolumnfalse} \vspace{0.6cm}

  ]

\renewcommand*\rmdefault{bch}\normalfont\upshape
\rmfamily
\section*{}
\vspace{-1cm}


\footnotetext{\textit{$^{a}$~Institute of Physical Engineering, Brno University of Technology, Technická 2896/2, 616 69 Brno, Czech Republic; E-mails: andrea.konecna@vutbr.cz, krapek@vutbr.cz}}
\footnotetext{\textit{$^{b}$~Central European Institute of Technology, Brno University of Technology, Purkyňova 123, 612 00 Brno, Czech Republic.}}

\footnotetext{\dag~Electronic Supplementary Information (ESI) available: details on the experimental procedures, fitting of core-loss EELS, raw low-loss EEL data, and computational details of ELNES calculations. See DOI: 00.0000/00000000.}



\section{Introduction}
Vanadium dioxide is a strongly-correlated material that can undergo a volatile phase transition between the low-temperature monoclinic insulating phase and the high-temperature rutile metallic phase \cite{Morosan2012_AdvMatReview,Budai2014_VO2Nature} by an applied thermal \cite{Nag2012_VO2-thermal}, electrical \cite{Wu2011_VO2-electric}, mechanical \cite{Aetukuri2013_strain}, or optical \cite{Lei2015_VO2-optic} biasing. In comparison to other phase-changing materials, VO$_2$ stands out for its ultrafast ($\approx 200$~fs) IMT with a convenient transition temperature ($\approx 67\,\degree$C) close to the room temperature \cite{Wegkamp2015_VO2ultrafast}, which can withstand millions of switching cycles without degradation \cite{Crunteanu2010_cycles}. All these properties suggest exciting applications of VO$_2$ in sensing \cite{Kim2007_temperature}, energy storage \cite{Khan2021_Batterisvo2}, switching \cite{Crunteanu2010_cycles,Yang2011_Switching,Dong2018_Switch}, data storage \cite{Lee2007_VO2-thermalprop2,Driscoll2009_UseMeta}, or active optical metamaterials \cite{Butakov2018_switchable,Howes2020_huygens,Kepic2021_ACS}. 

The properties of IMT in VO$_2$ can be strongly influenced by crystallographic defects resulting from a substrate lattice mismatch and specific fabrication conditions \cite{Aetukuri2013_strain}. IMT is also often dependent on VO$_2$'s grain size, strain, and film thickness \cite{Brassard2005grain,Wan2019_VO2-OnTheOpticalProperties,Jones2010_SNOM_microcrystals,Liu2013_snom_strain,Guo2011_wires_strain}. Further complexity stems from the existence of multiple stable stoichiometries of vanadium oxide which can coexist within one sample or functional structure\cite{Griffiths2003_stoichiometry,Lee1996_stoichiometry,Zhang2015_stoichiometry}. Many of the applications of VO$_2$ rely on nanostructuring, for which it is possible to use either top-down or bottom-up approaches \cite{Sohn2007_VO2nanowires,Joyeeta2008_Deposition,Li2014_VO2M,Ligmajer2018}. It is thus desirable to analyze IMT and stoichiometry of vanadium oxide locally.

The IMT is tightly connected with an abrupt change in electronic structure, which determines electric and optical properties. Conventional ways to characterize the IMT thus involve measurements of the electric conductivity \cite{Mitsuishi1967_phase} or optical response \cite{Wan2019_VO2-OnTheOpticalProperties,Moatti2019_electrical} as functions of applied temperature. However, these methods have limited spatial resolution and do not allow the analysis of the IMT at the level of single nanostructures or even grains. One possible solution is represented by scanning near-field optical microscopy (SNOM), which helped to identify local inhomogeneities in the IMT with a spatial resolution of $\sim 10^{1}$~nm \cite{Qazilbash2007_mott,Jones2010_SNOM_microcrystals,Liu2013_snom_strain,Ocallahan2015_inhomogeneity,Stinson2018_imaging}. Electron energy-loss spectroscopy (EELS) in a scanning transmission electron microscope (STEM) \cite{Egerton2011} is another suitable technique allowing for spatial resolution down to single atoms, which can also be used to correlate the spectroscopic information with high-resolution imaging \cite{Varela2004,Hage2020_Single-atomVibEELS}. Furthermore, EELS is particularly broadband and offers information on core-electron excitations with energies $\sim 10^2$~eV, optical excitations around $\sim 10^0$~eV and recently also vibrational excitations in $\sim 10^{-2}$~eV range \cite{Yang2022_ACSNano}.

EEL spectra in the core-loss region are dominated by excitations of electrons from core shells to unoccupied valence states. The spectral signatures are thus not only element-specific, but they also provide insights into local bonding, oxidation states of atoms \cite{Tan2012_EELSoxides}, and geometrical confinement \cite{Batson1993_core_loss_geometry}.  Several studies have already exploited EELS near-edge fine structure to identify different stoichiometries of  V$_x$O$_y$ samples \cite{Hebert2002,Su2003_highresEELS,Li2014_EELS-Francouzi,zhang2015wafer,Moatti2017_STEM_EELS,Zhang2017_wire_irradiation_EELS,Gauntt2009_MM_EELS,Zhou_interfacial_layers}. However, to the best of our knowledge, the studies using (S)TEM in combination with EELS were up to one exception \cite{Moatti2019_electrical} restricted to room temperature without the possibility of observing the changes when the material is switched to the high-temperature phase. Similarly, temperature-dependent low-loss STEM-EELS data that can be directly linked to optical properties are missing. In conducting materials, low-loss EELS is contributed by the excitation of plasmon resonances. This allows for a reliable distinguishing of conducting and insulating phases.

To gain insights into temperature-dependent properties of VO$_2$ at the microscopic level, we study a lamella with gradually decreasing thickness fabricated from a thick evaporated layer of \textit{presumably} VO$_2$ by focused ion beam (FIB) lithography. The lamella is attached to a commercial heating chip, which allows for external control of the lamella temperature. The lamella is then thoroughly studied by analytical methods available in TEM. We use high-resolution imaging and STEM with energy-dispersive X-ray spectroscopy (EDX) to analyze the lamella structure and elemental composition. The optical and electronic properties of the lamella are characterized by temperature-dependent STEM-EELS. We complement our experimental observations with ab-initio calculations of the core-loss probability. Low-loss probability is retrieved from the experimental dielectric function using a semi-classical formalism \cite{GarciadeAbajo2010_OptExc}.

\section{Methods}

\subsection{Preparation of thick VO$_2$ layer}
\new{A }550-nm-thick VO$_2$ layer was fabricated on a silicon substrate by evaporating VO$_2$ powder (Mateck) in an electron beam evaporator (Bestec, 8~kV, 32~mA, 1~Å/s) at room temperature and subsequent ex-situ annealing for 20 min at $550\,\degree$C in a vacuum furnace under 10 sccm of O$_2$ flow.

\subsection{Preparation of the lamella}
Lamellas for TEM were fabricated using a gallium FIB Tescan LYRA3. 
The final polishing \new{and thinning of the V-shaped lamella denoted in the following as ``IS''} was performed with \dlt{imaging at}\new{ a} low voltage (5 kV) and low current (40 pA) \new{FIB} \dlt{of Ga+ ions} for approximately 25 seconds at a \new{low} glancing angle \new{(the ion beam is nearly parallel with the surface of the lamella)} to reduce the beam-damaged surface area \cite{Mayer2007_temSample}. The positioning of the lamella on the heating chip is shown in detail in Figure~S1 of the Supplementary Information.

\subsection{Transmission electron microscopy}
TEM analysis was performed on TEM FEI Titan equipped with a Super-X spectrometer for EDX, a monochromator, and a GIF-Quantum spectrometer for EELS. The heating experiment was performed using in-situ holder Fusion Select and the corresponding thermal chip by Protochips \cite{protochips}. \new{To avoid excessive electron-beam-induced damage to the sample, we have utilized as low electron currents as possible while keeping a good signal-to-noise ratio. Further, we performed each step of the analysis in a new spot on the sample, so far unexposed. On the other hand, the distance between the spots was minimized to reduce the impact of the sample's inhomogeneity.}

\subsubsection{STEM-EDX}
EDX measurement was performed in the scanning regime with the energy of primary electrons set to 300~keV. The probe current was set to approximately $600$~pA. We acquired and integrated 100 spectrum images with a pixel size of 0.34~nm and a dwell time of 2~$\upmu$s. The spectrometer dispersion range was set to 20~keV. Chemical composition was evaluated in Velox software using the parabolic background model, absorption correction, and Schreiber-Wims ionization cross-section model (a semi-empirical model recommended for metal oxides). For the quantification of the vanadium dioxide layer stoichiometry, we used V \textit{K}, O \textit{K}, and Ga \textit{K} edges.

\subsubsection{STEM-EELS core-loss}
Core-loss EELS was performed in the monochromated scanning regime with the energy of primary electrons set to 300~keV. We used the following parameter settings: The probe current of around $200$~pA, the convergence semi-angle of 10~mrad, the collection semi-angle of 18.4~mrad, and the spectrometer dispersion of 0.25~eV/px. The full-width at half-maximum of the zero-loss peak (determining the energy resolution) read 0.2~eV. We used the dualEELS option, which allows us to record both the low-loss and core-loss EELS spectra simultaneously, resulting in the correct calibration of the energy loss axis to absolute values \cite{Scott2008_dualEELS}. The spectrum line scans were acquired with a length of 100 pixels, a pixel size of 1.44~nm, and an exposure time of 0.1~ms for the low-loss part and 1~s for the core-loss part, respectively. The low-loss part of the spectrum was further used to determine the thickness profiles in Figure~\ref{fig:Figure_2}(c) in the units of \new{inelastic mean free path (}IMFP\new{)}. 

\subsubsection{STEM-EELS low-loss}
Low-loss EELS was performed in the monochromated scanning regime with the primary electron energy of 120~kV. The reduced value of the primary electron energy was used to achieve a better signal-to-background ratio \cite{Horak2020_ExpConditions} and to suppress relativistic losses like the Čerenkov radiation \cite{Horak2015_Cerenkov}. We used the following parameter settings: the probe current of approximately $100$~pA, the convergence semi-angle of 10~mrad, and the collection semi-angle of 11.4~mrad, the spectrometer dispersion of 0.1 eV/pixel. The full-width at half-maximum of the zero-loss peak (determining the energy resolution) read 0.14 eV. We summed 100 spectra recorded over a $10\times 10$~pixel spectrum image with a pixel size of 1~nm and an exposure time of 0.8~ms. The zero-loss peak subtraction was performed by a background power-law fitting respecting the shape of the pre-recorded zero-loss peak at vacuum.

\section{Results and Discussion}
\subsection{Sample Fabrication and Pre-Characterization}

\begin{figure*}
  \includegraphics[width=0.95\linewidth]{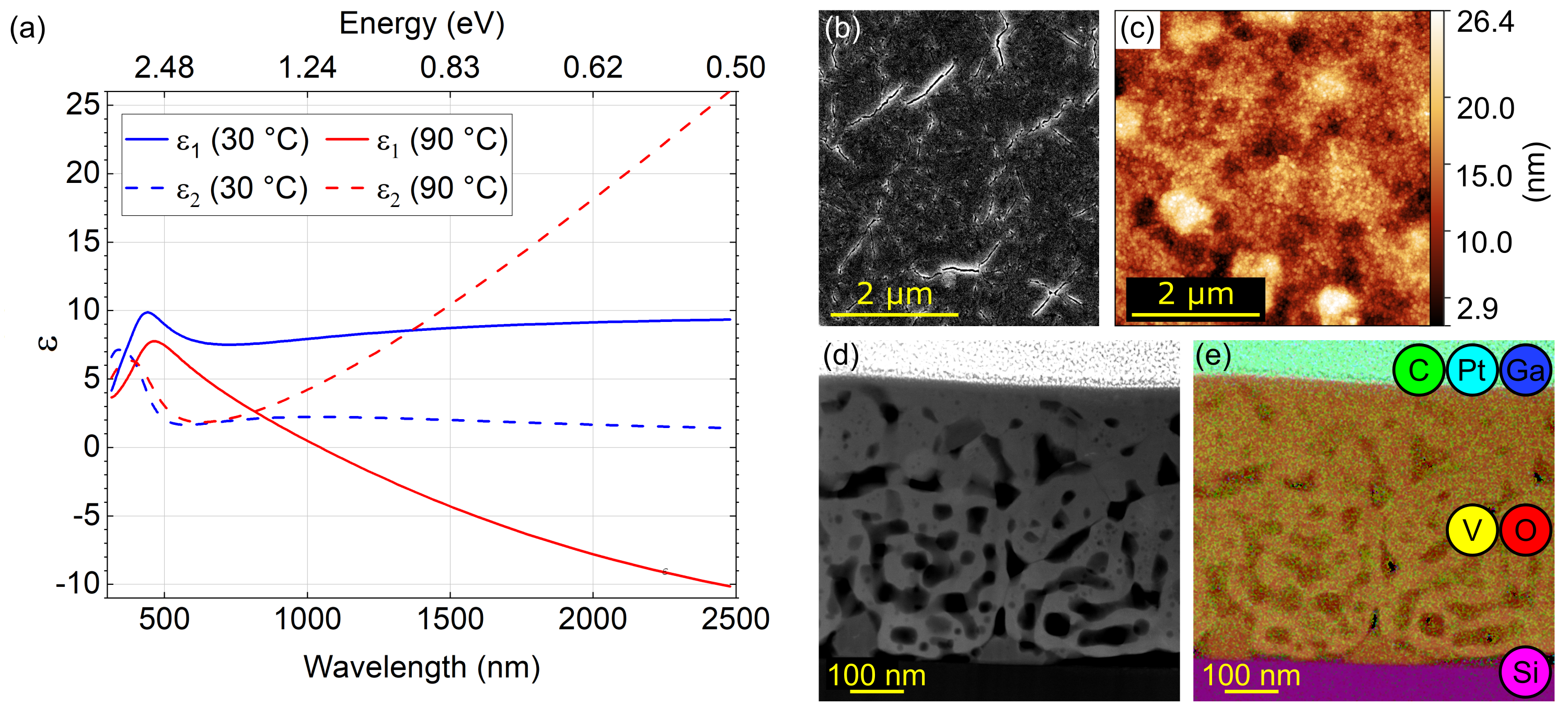}
  \centering
  \caption{Pre-characterization of the VO$_2$ thin film. 
  (a)~Optical properties below (blue) and above (red) IMT temperature 
  ($\approx$\,67\,$\degree$C).
  (b)~Illustrative SEM image of the surface of the sample.
  (c)~AFM image of the surface from a different place than (b).
  (d)~STEM-HAADF image of a thin lamella ``S'' displaying a porous character of VO$_2$ layer. 
  (e)~STEM-HAADF image with overlayed EDX maps from the same place as (d) visualizing the presence of individual elements detected in the sample: Si substrate at the bottom (pink); V (yellow) and O (red) in the middle; a mixture of C (green), Pt (light blue), and Ga (dark blue) as components of the Pt protective coating at the top.}
  \label{fig:Figure_1}
\end{figure*}

A layer of \textit{presumably} vanadium dioxide (VO$_2$) with a thickness of 550\,nm was prepared by electron beam evaporation and subsequent ex-situ annealing (see Experimental Section). We first performed macroscopic characterization of the sample by spectroscopic ellipsometry to verify that it indeed corresponds to VO$_2$ and that it exhibits the desired optical properties and their switching across the IMT temperature as reported in literature \cite{Wan2019_VO2-OnTheOpticalProperties,Kana2011_elips}. \textbf{Figure 1}(a) shows real (solid lines; $\varepsilon_1$) and imaginary (dashed lines; $\varepsilon_2$) parts of dielectric function at temperatures below (30\,\degree C, blue) and above (90\,\degree C, red) the IMT temperature. The high-temperature phase is metallic, with a negative $\varepsilon_1$ function above the wavelength of 1100 nm, while the low-temperature phase is insulating, with positive $\varepsilon_1$ in the whole inspected spectral region. The onset of interband electronic absorption is observed below about 500~nm, manifested by a peak in $\varepsilon_2$. Above 500 nm, $\varepsilon_2$ increases as $\sim\lambda$ in the metallic phase as a consequence of the Drude-like response and remains low in the insulating phase. The IMT in VO$_2$ associated with the observed qualitative change in its optical properties gives rise to a possible application of VO$_2$ structures in active photonics \cite{Kepic2021_ACS}.

We further characterized the homogeneity of the layer surface by a scanning electron microscope (SEM) and an atomic force microscope (AFM) as shown in Figure~\ref{fig:Figure_1}(b,c), respectively. The SEM image shows small cracks throughout the whole surface of the layer. The AFM image obtained at a different place on the surface also shows roughness and apparent pores \new{ in line with previous reports~\cite{MARVEL2015217}}. 
The presence of inhomogeneities observable already at the surface triggered the need for a cross-sectional structural and chemical analysis using TEM. 

We prepared a standard TEM lamella (further labelled as ``S'' which \dlt{stays}\new{stands} for standard) by FIB. A high-angle annular dark-field (HAADF) image of the $\sim 120$~nm thick lamella (representing a cross-section of the original VO$_2$ layer) is shown in Figure~\ref{fig:Figure_1}(d). The transverse profile of the layer exhibits pores and cracks noticeable also in the SEM and AFM images. The TEM image overlayed by a map of the elemental composition obtained by EDX is shown in Figure~\ref{fig:Figure_1}(e). From the bottom up we identify the silicon (pink) substrate,  vanadium (yellow) and oxygen (red) forming the layer itself and a protective layer -- a mixture of Pt (cyan), C (green), and minor contribution of implanted Ga (dark blue) prepared by focused ion beam induced deposition (FIBID). The EDX analysis of the VO$_2$ layer reveals the elemental composition reading $32 \pm 6$ atomic percent of vanadium, $68 \pm 8$ atomic percent of oxygen, and $0.3 \pm 0.1$ atomic percent of gallium (only negligible contamination). The mean stoichiometry VO$_{2.1 \pm 0.4}$ determined by EDX is consistent with the expected VO$_2$ composition of the layer. \new{Within porous areas of the sample [dark areas in Fig.~\ref{fig:Figure_1}(d)] we found increased oxygen amount with the elemental rates of $(28 \pm 5)\,\%$ of vanadium and $(72 \pm 9)\,\%$ of oxygen. As we discuss in Section~S7 of the Supporting Information, the surplus oxygen corresponds to the contamination rather than the apparent mean stoichiometry V$_2$O$_{5.1\pm 1.1}$ corresponding to V$_2$O$_5$.}

To perform in-situ analysis of IMT in the TEM, another lamella (further labeled as lamella ``IS'' standing for in-situ) was prepared and attached to a TEM heating chip (Protochips \cite{protochips}; see Figure S1 of the Supporting Information). It consisted of a thick support and a thinned layer with gradually decreasing thickness dedicated for EELS measurements. Figure~\ref{fig:Figure_2}(a) shows a STEM annular dark-field (ADF) image with an overview of the lamella ``IS''. We further performed high-resolution imaging of the lamella and confirmed its polycrystallinity with random orientation of grains as shown in Figure S2 of the Supporting Information. The yellow frame in Figure~\ref{fig:Figure_2}(a) indicates the thinnest region of the lamella selected for further analysis.

\subsection{Core-Loss EELS}

\begin{figure*}
\centering
  \includegraphics[width=0.75\linewidth]{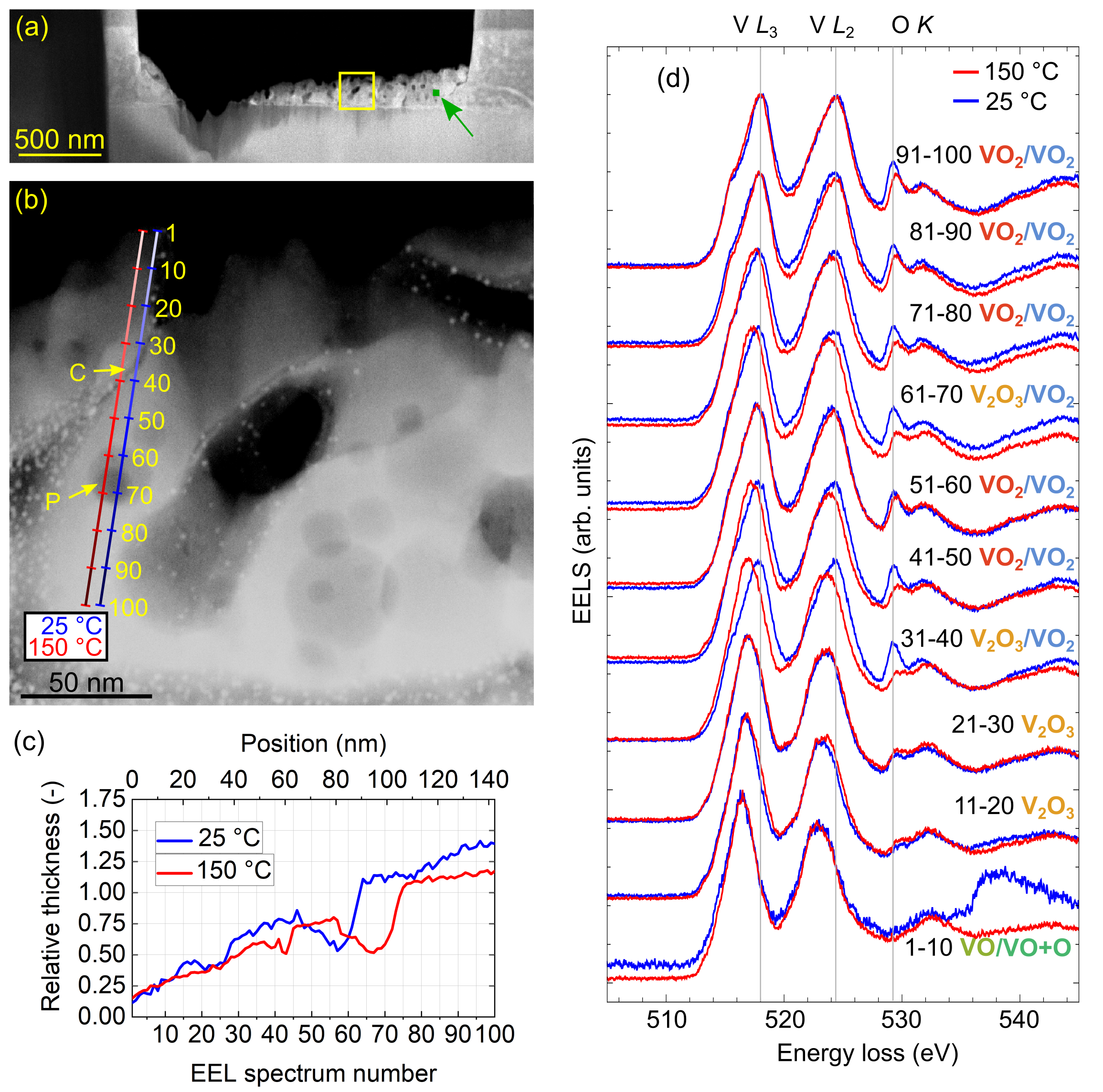}
  \caption{EEL\dlt{S} spectrum line scans at nominal temperatures 25$\,\degree$C and 150$\,\degree$C. 
  (a) STEM-ADF image of the lamella ``IS''. Yellow-framed square marks the region of interest for the spectrum line scans, whereas the small green square, also denoted by the green arrow, will be used later for the low-loss EELS experiment. (b) Detail of the yellow-marked area. The blue line corresponds to a desired spectrum line scan at 25$\,\degree$C, and the red line corresponds to a line spectrum image at 150$\,\degree$C. Both spectrum line scans comprise 100 individual EELS measurements. Yellow arrows mark the crack (C) and pore (P).
  (c) Relative thickness in the units of IMFP, which varies around 130~nm for different V$_x$O$_y$ stoichiometries extracted from EELS. Note that the line scan at 25$\,\degree$C slightly drifted from the desired position displayed in (b).
  (d) 2D waterfall graph with measured core-loss EEL spectra at both temperatures. Each spectrum averages ten subsequent EEL spectra as marked next to them. Vertical gray lines are to guide the eyes.}
  \label{fig:Figure_2}
\end{figure*}

Our VO$_2$ layer exhibits structural inhomogeneities, which might be associated with chemical inhomogeneities and irregularities in the switching from the insulating to the metallic phase. To explore this possible correlation, we analyze EELS of the lamella "IS" in the core-loss region. In Figure~\ref{fig:Figure_2}(b), we show the target line scans along which we acquire $2\times 100$ spectra with regular spacing. We perform the spectrum line scan close to room temperature (blue line) and above IMT temperature (red line). The accuracy of the nominal temperature at the chip is reported to be $5\,\%$ \dlt{\cite{Idrobo2018_temperature}, but as contacts between the lamella and the heating chip might not be perfect and cause losses,}\new{for a planar specimen positioned directly on the chip~\cite{Idrobo2018_temperature}. However, the lamella is mostly free-standing and only connected to the heating chip by narrow pillars (see Fig.~S1 in the Supporting Information) with presumably mediocre heat conductivity. Therefore, we assumed a larger discrepancy between the nominal and real temperature, and} we heated the chip to 150$\,\degree$C to ensure we were certainly above the IMT temperature. We also note that the designated line scan positions, especially the one for the measurement at 25$\,\degree$C, slightly differ from the actual measurements due to sample drift. The thickness profiles extracted from EELS measurements in the units of the electron \dlt{inelastic mean free path (}IMFP\dlt{)} \new{using the log-ratio method~\cite{Egerton2011,https://doi.org/10.1002/jemt.1060080206}} are shown in Figure~\ref{fig:Figure_2}(c). The IMFP for our experimental parameters and possibly occurring different V$_x$O$_y$ stoichiometries varies around 130~nm (134~nm for VO$_2$, 131~nm for V$_2$O$_3$ and 127~nm for VO based on the Mean free path estimator \cite{Mitchell2006_IMFP} with model from Ref.~\cite{Malis1988_thickness}). Both line scans correspond to gradually increasing thickness with several valleys associated with the presence of the pores.

To facilitate visualization of the spectra and preliminary analysis, we binned the spectra into twenty groups by ten, and represented each bin by the average spectrum, which was further normalized by its maximum. The spectra shown in the waterfall plot in Figure~\ref{fig:Figure_2}(d) globally exhibit up to five prominent distinct peaks. Notably, we can observe energy shifts and intensity changes in the peaks appearing in the spectra averaged over different beam positions. An additional feature, a pre-peak (a ``shoulder") of the first peak, is emerging in the spectra recorded in thicker areas. However, the changes in subsequent averaged spectra are rather abrupt, suggesting that a finer analysis is desirable.

\begin{figure*}
\centering
  \includegraphics[width=\linewidth]{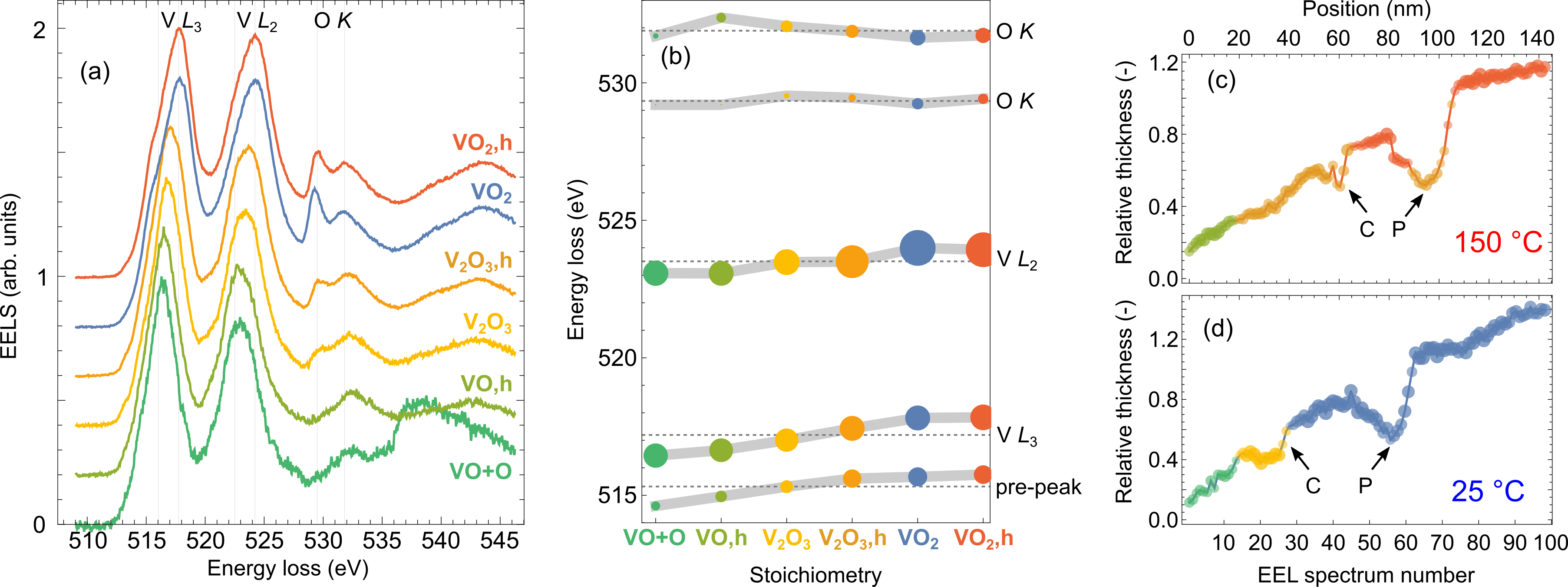}
  \caption{Cluster\dlt{ing} analysis of the spectrum line scans from Figure~\ref{fig:Figure_2}. 
  \dlt{2}\new{1}00 spectra recorded within \dlt{the}\new{each} spectrum line scan\dlt{s} are sorted into \dlt{four}\new{three} groups (clusters) according to similarities. \dlt{Averaged normalized spectra falling}\new{The normalized averages of the spectra belonging} into the individual clusters \new{(cluster representants)} are shown in (a). Fitting of the spectra in (a) by a sum of Gaussians yields peak energies plotted in (b), where the sizes of the dots are proportional to the \new{square root of} spectral areas of the peaks. Line profiles in (c,d) in the units of mean free path are overlaid by color-coded dots denoting affiliation of the spectrum recorded at the corresponding position to one of the clusters. The dots' size and opacity denote closeness to each \dlt{cluster's averaged spectrum}\new{cluster representant}. The color-coding corresponds to (a). (c) is obtained for the measurement at 150$\,\degree$C and (d) for 25$\,\degree$C. Arrows mark the position of the crack (C) and the pore (P).}
  \label{fig:Figure_3}
\end{figure*}

We utilize cluster analysis for a refined sorting of the acquired spectrum line scans. \dlt{We first group the spectra acquired at both temperatures (200 spectra in total) and then sort them} \new{We sort each dataset (spectra acquired at room and elevated temperature, respectively)} into clusters using the ``k-means" clustering approach as implemented in Mathematica 12 \cite{mathematica}. Inspecting the results of several clustering procedures (with the number of clusters \new{for each dataset} varied between \new{2 and 4} \dlt{3 and 5; and with low-temperature and high-temperature spectra processed separately or jointly}), we identified \dlt{four} \new{three} clusters as optimum -- all clusters were sufficiently populated and represented physically distinct spectra. \dlt{We further notice that one of the clusters can be sorted into two sub-clusters based on spectral behavior at energies above 535~eV.} Averaged spectra corresponding to the \new{six} \dlt{five} acquired clusters, referred to as cluster representants, are plotted in Figure~\ref{fig:Figure_3}(a). \dlt{The two green spectra emerge from the sub-clustering.}

The cluster representants exhibit multiple peaks, which can be assigned to transitions from core to unoccupied valence states of vanadium \cite{Gloter2001} and oxygen \cite{Hebert2002}: i) V \textit{L}$_3$ peak (\dlt{518--519} \new{516--518} eV; transitions from 2\textit{p}$_{3/2}$ to unoccupied 3\textit{d} states of V), ii) V \textit{L}$_2$ peak (\dlt{521--522} \new{523--524} eV; transitions from 2\textit{p}$_{1/2}$ to unoccupied 3\textit{d} states of V), and iii) up to two O \textit{K} peaks (529--532 eV; transitions of O electrons from 1\textit{s} to hybridized O 2\textit{p} – V 3\textit{d} states, which are of $t_{2g}$ and $e_g$ symmetry). Spectral features observable above $\sim 535$~eV arise due to transitions to O 2\textit{p} and V 4\textit{s} mixed states \cite{Hebert2002}, but can be influenced by the presence of unbound oxygen or other oxygen compounds (\textit{e.g.}, due to contamination~\cite{doi:10.1021/acs.chemrev.9b00439}). 

We further fit the cluster representants by a sum of \dlt{eight} \new{seven} Gaussian peaks (for details on fitting, see Figure~S3 of the Supporting Information)\dlt{, whose energies} \new{. Energies of five of the fitted peaks corresponding to spectral features of our interest} are plotted in Figure~\ref{fig:Figure_3}(b). Notably, energies and relative strengths of V and O peaks strongly depend on the local oxidation state. The energies of V \textit{L}$_3$ and \textit{L}$_2$ peaks increase with a higher oxidation state of V, and importantly, we observe the emergence of a pre-peak, which has been attributed as a clear signature of crystalline VO$_2$\dlt{.} \cite{Gloter2001,Li2014_EELS-Francouzi}\new{.} Another clear signature is the increasing intensity of the first O \textit{K} peak at $\sim 529$~eV (V 3\textit{d} $t_{2g}$ bands) \cite{Hebert2002}. Such analysis allows for the precise assignment of our clusters with different V$_x$O$_y$ stoichiometries. We find that the first cluster \new{in each dataset} can be attributed to VO \new{due to the missing first O \textit{K} peak}. \new{However, we notice a difference between the cluster representants in the region around 539~eV. A broad peak associated with} \dlt{The sub-clustering then results in distinguishing VO and VO$+$O, where the latter corresponds to areas with} the O in other compounds or free O\new{\cite{doi:10.1021/acs.chemrev.9b00439}} \new{appears for the case of the measurement at room temperature. We thus denote the cluster representant as ``VO+O''.} \dlt{, yielding an increase of a broad peak centred around 539~eV [green spectra in Figure~\ref{fig:Figure_3}(a)]}. We then identify V$_2$O$_3$ (orange \new{and yellow cluster representants} \dlt{spectrum}) with \dlt{nearly equal intensities of} \new{two emergent} O \textit{K} peaks \new{of comparable intensities}. The last two \dlt{clusters} \new{cluster representants obtained for room and elevated temperature} (blue and red spectra\new{, respectively}) exhibit a clear pre-peak around 515~eV and an intense first O \textit{K} peak, which for both yields the VO$_2$ assignment. However, a closer inspection reveals a slight energy shift \dlt{between the V peaks} and a different intensity ratio of O peaks between these two clusters. \new{Interestingly, we did not identify any signature of V$_2$O$_5$, although it is rather common in nominally VO$_2$ thin films~\cite{MARVEL2015217}. EELS of V$_2$O$_5$ is typical by much higher contribution of the O \textit{K} peaks compared to V \textit{L} peaks~\cite{Su2001_electronReduction}.}

It is interesting to correlate the cluster index of each spectrum (i.e., to what cluster an individual spectrum belongs) with the parameters at which the spectrum was taken: the position at the sample, the local thickness of the sample, and the temperature. We visualize this in Figure~\ref{fig:Figure_3}(c,d). We represent each spectrum as a point in the sample-position--local-thickness coordinates, the color of the point represents the cluster index, and each panel corresponds to one of the two temperatures. \new{Apparently, there is a strong correlation between the cluster index and the local thickness. On the other hand, as long as the thickness is identical, we do not observe significant differences between the crack, the pore and the homogeneous part of the lamella. Another significant parameter is the temperature. For each of the identified stoichiometries, the cluster representants at the room and the high temperature differ significantly, as shown in Fig.~\ref{fig:Figure_3}(b). We note that this difference is a mere indication of thermally dependent properties of all considered vanadium oxides (VO, V$_2$O$_3$ and VO$_2$), but it does not provide an evidence of the phase transition without further analysis.}
\dlt{For clusters 1a/b and 2 corresponding to VO/VO$+$O and V$_2$O$_3$, respectively, the only discriminating parameter is the sample thickness, and the spectra taken at both temperatures at a specific thickness range are assigned to the same cluster. Notably, for clusters 3 and 4 containing spectra from thicker parts of the sample, the temperature becomes a discriminating parameter -- cluster 3 contains only low-temperature spectra (77 in total), and cluster 4 contains only high-temperature spectra (36 in total), with only two exceptions. This allows us to state already at the phenomenological level that the thicker part of the sample assigned to clusters 3 and 4 exhibits a pronounced thermal dependence of its properties (possibly IMT) while the thinner parts assigned to clusters 1 and 2 show no thermal dependence making the occurrence of IMT improbable.}

The cluster analysis yields \dlt{two conclusions: i)}\new{a following conclusion:} With decreasing lamella thickness, we observe a gradual reduction of vanadium from VO$_2$ to V$_2$O$_3$, and VO/VO+O. The part of the lamella with the reduced oxides is rather large, with a thickness up to 0.6\,--\,0.7~IMFP or 80\,--\,90~nm and a width (measured from the lamella edge) up to 60~nm [cmp. Figure~\ref{fig:Figure_3}(c,d)]. This suggests difficulties in preparing ultrathin VO$_2$ nanostructures, as the reduced oxides do not exhibit the desired switching of optical properties. On the other hand, a previous study reported considerably thinner (about 2~nm) diffuse layers of reduced vanadium oxides at the boundaries of individual VO$_2$ grains \cite{Li2014_EELS-Francouzi}, suggesting that with optimized fabrication protocols, the detrimental impact of the reduced oxides can be relieved. \dlt{ii) The spectra from the thickest areas acquired at 25$\,\degree$C and 150$\,\degree$C are unambiguously sorted into the two substantially different VO$_2$ clusters [blue vs. red spectrum in Figure~\ref{fig:Figure_3}(a)], which suggests that the observed spectral and intensity shifts between the two VO$_2$ clusters could be attributed to the phase change. This assignment is further corroborated by the ab-initio calculations discussed below.} 

Finally, we discuss the role of the structural inhomogeneity in the local properties of the lamella. The line scans shown in Figure~\ref{fig:Figure_2}(b) cross two pronounced inhomogeneities: a narrow crack at the targeted spectrum indices of 35 (\dlt{low}\new{room} temperature, blue line) and 40 (high temperature, red line) and a broad deep pore at the targeted spectrum index range of 60 to 70 for both temperatures. Both the crack and the pore are visible as depressions in the thickness profiles shown in Figure~\ref{fig:Figure_2}(c). Here, the real spectrum indices slightly differ from the targeted ones due to sample drift, which resulted in an offset of 10 between the targeted and real spectrum indices for the \dlt{low}\new{room} temperature. The crack is found at the spectrum index of 22 (\dlt{low}\new{room} temperature) and 43 (high temperature), and the pore is located in the spectrum index ranges of 46 to 64 (\dlt{low}\new{room} temperature) and 58 to 75 (high temperature).

For the crack at both temperatures and for the pore at the \dlt{low}\new{room} temperature we observe no modification of the core-loss EEL spectrum in comparison to the areas of the lamella adjacent to the inhomogeneity, neither by visual inspection [Figure~\ref{fig:Figure_2}(d)] nor by cluster analysis represented by the spectrum cluster index [Figure~\ref{fig:Figure_3}(d)]. On the other hand, the core-loss EEL spectrum is modified for the deepest part of the pore at the high temperature, where the stoichiometry changes from VO$_2$ to V$_2$O$_3$ [see Figure~\ref{fig:Figure_3}(c)]. A closer inspection suggests that this modification is related to a reduced thickness of the lamella within the pore. In other words, the singular parameter determining the local electronic and optical properties of the lamella manifested in core-loss EELS is its thickness. The structural inhomogeneities (cracks and pores) themselves do not impose any observable modifications (e.g. due to their elastic or plastic field\new{, different local composition, stoichiometry,} or local charge) unless they are accompanied by a modified \new{local} thickness. \new{As an important consequence, the chemical composition and stoichiometry of the vanadium oxide inside cracks and pores do not differ from those far from the cracks and pores (for the same local thickness). This also means that the presence of cracks and pores is not detrimental to the optical properties of the vanadium dioxide layer (while we cannot exclude an impact on its mechanical or transport properties).}

We note that the area of the lamella identified as the crack is present in the region of V$_2$O$_3$ at the interface with VO$_2$. Further investigation is required to determine whether this correlation is causal or coincidental, and possibly determine its mechanism.

\subsection{Variation of Core-Loss and Low-Loss EELS with Temperature}
\begin{figure*}
\centering
  \includegraphics[width=\linewidth]{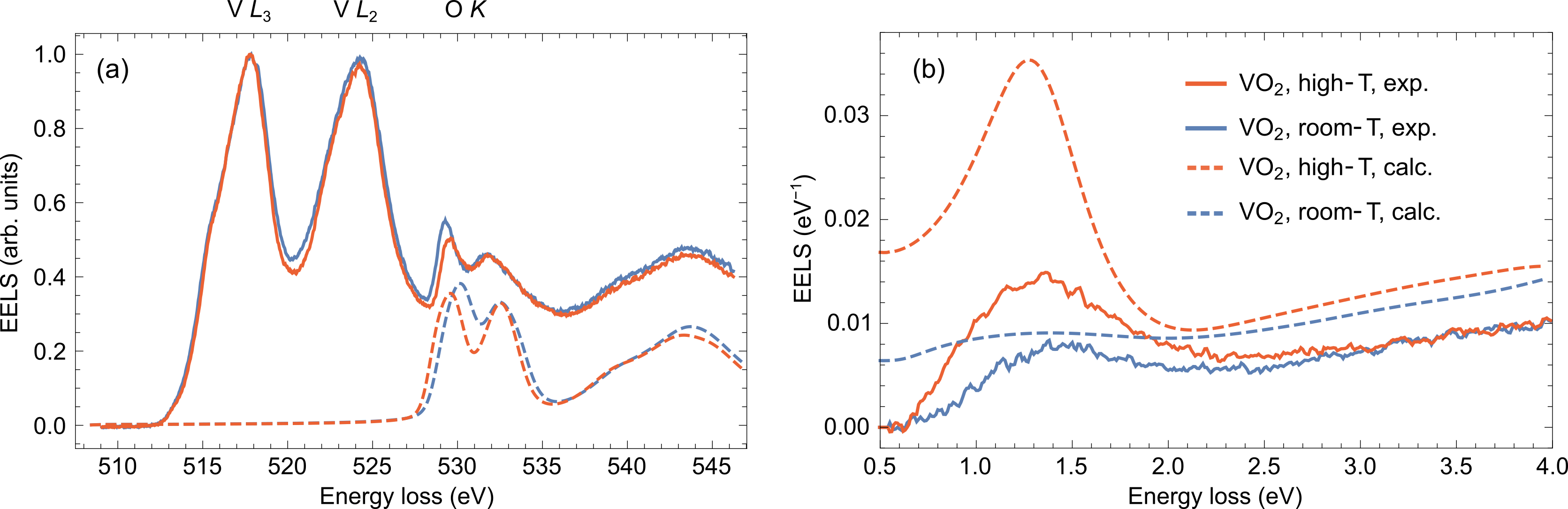}
  \caption{(a) Comparison of the averaged clustered spectra (cluster representants) assigned as VO$_2$ obtained at 25$\,\degree$C (solid blue line), and 150$\,\degree$C (solid red line) with ab-initio-calculated O \textit{K} peaks corresponding to the monoclinic phase (dashed blue line) and rutile phase (dashed red line). The calculated spectra are both offset by \dlt{527.8}\new{528.5}~eV to get an approximate alignment with the experimental peak positions. (b) Comparison of the experimentally acquired low-loss EELS at 25$\,\degree$C (solid blue line), and 150$\,\degree$C (solid red line) from the area marked by the \new{small} green square \new{and arrow} in Figure~\ref{fig:Figure_2}(a) compared with calculations using a classical dielectric formalism (dashed blue and red lines).}
  \label{fig:Figure_4}
\end{figure*}

\dlt{For further verification that the temperature-dependent changes in our core-loss EELS measurements are due to the IMT of VO$_2$, we perform ab-initio calculations.}\new{To verify that the temperature-dependent changes in our core-loss EELS measurements of VO$_2$ are due to the IMT, we perform ab-initio calculations.} We use the TELNES3 package, the extension of the Wien2k all-electron DFT code \cite{blaha_wien2k_2020}, to simulate electron loss near-edge structure (ELNES) of the O~\textit{K}-edge. More details about the calculation setup are given in the Supporting Information.

In Figure~\ref{fig:Figure_4}(a), we present a comparison between the cluster representants of core loss EELS obtained in the thick areas of the sample at 25$\,\degree$C and 150$\,\degree$C (solid lines) together with the calculated spectra (dashed lines; after convolution with 1-eV Gaussian) concerning the O~\textit{K} excitations.
The ab-initio theory predicts changes in the positions and relative strengths of the convolved O~\textit{K} peaks, \new{the latter} in \dlt{perfect}\new{qualitative} agreement with our experimental observation. \dlt{The relative shift of $\sim 0.2$~eV between the high-temperature and low-temperature experimental spectra on the energy axis is caused by the band gap opening in the monoclinic phase. Our calculations show 0.1~eV gap.} In line with the previous work of Hebert et al.\cite{Hebert2002}, the relative intensity of the O~\textit{K} peaks is proportional to the number of O(p) states in the \textit{t$_{2g}$} band ($\pi$-bonds\new{; first O \textit{K} peak}) and in the \textit{$e_g$} band ($\sigma$-bonds\new{; second O \textit{K} peak}). Indeed, the ratio of O(p) states in respective bands changes from \dlt{77.4\% to 80.7\%}\new{$80.7\,\%$ to $77.4\,\%$} after IMT, which explains a relative \dlt{increase}\new{decrease} of the first O~\textit{K} peak \new{with respect to the second O~\textit{K} peak at elevated temperature}. 

Modifications in the band structure across the IMT are also tightly connected with changes in valence-electron excitations, which can be probed by EELS in the low-loss region. In the case of VO$_2$, the IMT could be clearly identified based on the emergence of plasmons in the rutile (metallic) phase. Plasmons in VO$_2$ nanostructures were previously detected only with high-resolution EELS \cite{Felde1997_HREELS_plasmon_VO2}, where slow electrons in a broad beam were used to probe sample's surface. However, to the best of our knowledge, low-loss STEM-EELS of nanostructured VO$_2$ samples probed across the IMT have been missing.

Unfortunately, the long time necessary to acquire the core-loss EELS prevented us from performing the low-loss EELS experiment in the previously analysed area due to accumulated contamination and beam-induced damage. We therefore collected the low-loss EEL spectra in the region close by, denoted by the green square in Figure~\ref{fig:Figure_2}(a). This region is sufficiently thick ($\sim 300$~nm based on STEM-HAADF images of the lamella and a comparison of intensities with areas of known thickness) without too many pores, so it contains dominantly VO$_2$. To avoid beam-induced damage, we collected a spectral image with $10\times 10$ pixels with regular steps of $\sim 1$~nm and averaged them. We further fitted and subtracted the zero-loss peak to obtain the results in Figure~\ref{fig:Figure_4}(b) (solid lines; see raw spectra in Figure~S4 of the Supporting Information). The experiment was again performed below and above the IMT temperature, at 25$\,\degree$C (blue) and 150$\,\degree$C (red), respectively.

The experimental low-loss EEL spectra in Figure~\ref{fig:Figure_4}(b) exhibit a broad peak between $1.0-1.5$~eV, which grows in intensity when the sample is heated up. To interpret this behavior, we perform semi-analytical calculations within the framework of classical electrodynamics. To that end, we consider the problem of a line current (representing a focused electron beam) traversing a thin film characterized by a dielectric function from Figure~\ref{fig:Figure_1}(a) and solve for the induced electric field entering the formula for electron energy-loss probability \cite{GarciadeAbajo2010_OptExc}. The spectrum calculated for VO$_2$ in the monoclinic (insulating) phase (dashed blue line) shows only a faint broad peak centered at $\sim 1.25$~eV corresponding to weak absorption contribution, a small portion of radiative losses and an increasing background above $2.5$~eV due to higher-energy interband transitions. On the other hand, the intense peak around $1.3$~eV appears in the calculation for the rutile (metallic) phase (dashed red line), which can be assigned to bulk plasmon and surface plasmon polariton excitations. These theoretical results qualitatively support the experimental observation and confirm that the spectral change could be interpreted as IMT-induced. However, the experimentally measured loss probability is less intense, which is particularly apparent in the plasmonic peak in the spectrum recorded at 150$\,\degree$C. Such quantitative discrepancy can happen due to differences in the actual and model dielectric responses of the sampled area, or partially because of the background subtraction procedure, where zero signal is assumed below 0.5~eV. Multiple scattering, which is not included in the theory, could also play some role.

\section{Conclusions}
We have studied the insulating and metallic phases of nanostructured vanadium dioxide (VO$_2$) using analytical electron microscopy. We have focused on ultrathin lamellas fabricated from a 550~nm thick VO$_2$ film. With the help of STEM-EELS, we have revealed that the lamellas show both structural and stoichiometric inhomogeneities at the microscopic level. Results from the core-loss EELS enhanced by statistical analysis helped to correlate the stoichiometry with the thickness of the lamella, finding vanadium oxides with a reduced oxidation number (V$_2$O$_3$ and VO) at thinner parts of the lamella (with the thickness of 60~nm or less). This represents a challenge for preparing ultrathin VO$_2$ nanostructures.

The principal contribution of our study is the observation of temperature-induced changes in both core-loss and low-loss EELS in thicker regions of the lamella, which were identified to contain dominantly VO$_2$. Changes in the core-loss spectra were analyzed with the help of ab-initio calculations that were used to relate the fine structure of the oxygen \textit{K} peaks with the modifications in the band structure during the IMT. In the low-loss spectra, the most significant evidence of IMT is the emergence of a strong plasmonic peak in the metallic phase.

This work establishes the STEM-EELS with in-situ heating as a suitable technique to probe the IMT in VO$_2$, and potentially in other materials. We foresee that STEM-EELS could also be combined with in-situ diffraction studies to correlate the spectroscopic results with crystal structure and orientation to further enhance the analysis of the IMT. 


\section*{Conflicts of interest}
There are no conflicts to declare.

\section*{Acknowledgements}
This research was supported by the Czech Science Foundation (project No.~22-04859S) and MEYS CR under the project CzechNanoLab (LM2023051, 2023–2026). JP acknowledges the funding from IMPROVE V, CZ.02.01.01/00/22\_010/0002552.



\balance


\bibliography{references} 
\bibliographystyle{rsc} 

\end{document}